\documentclass[12pt,a4paper]{article}

\usepackage{amsmath}
\usepackage{amsfonts}
\usepackage{tikz}
\usepackage{pgfplots}

\title{ \textbf{Investment strategies based on forecasts are (almost) useless} }
\author{Michael Weba}

\date{}

\newcommand{\E}{\ensuremath{ \mathsf{E} }}
\newcommand{\A}{\ensuremath{ \mathcal{A} }}
\newcommand{\B}{\ensuremath{ \mathcal{B} }}
\newcommand{\F}{\ensuremath{ \mathcal{F} }}
\newcommand{\bE}  {\ensuremath
   { \mathsf{E}(p_S|\mathcal{B})  }  }

\begin{document} 

\maketitle

\large

\textbf{Abstract}. Several studies on portfolio construction reveal that sensible strategies essentially yield the same results as their nonsensical inverted counterparts; moreover, random portfolios managed by Malkiel's dart-throwing monkey would outperform the cap-weighted benchmark index. Forecasting the future development of stock returns is an important aspect of portfolio assessment. Similar to the ostensible arbitrariness of portfolio selection methods, it is shown that there is no substantial difference between the performances of ``best'' and ``trivial'' forecasts - even under euphemistic model assumptions on the underlying price dynamics. A certain significance of a predictor is found only in the following special case: the best linear unbiased forecast is used, the planning horizon is small, and a critical relation is not satisfied. \\


\normalsize

\vspace{2cm}

\textbf{Address:} \\
Michael Weba \\
Goethe University Frankfurt \\
Department of Economics and Business \\
Theodor-W.-Adorno-Platz 4 \\
60629 Frankfurt am Main, Germany \\

e-mail: weba@wiwi.uni-frankfurt.de

\newpage

\begin{large}

\section{Introduction}
B. Malkiel [2007] stated that ``a blindfolded monkey throwing darts at a newspaper's financial pages could select a portfolio that would do just as well as one carefully selected by experts''. In this vein, Metcalf and Malkiel [1994] presented a statistical analysis of thirty Wall Street Journal dartboard contests. Various tests were applied to compare US traded equity recommendations given by experts with the random selection of darts. The authors concluded that there is strong support for the hypothesis that experts cannot beat the market consistently and that stock picking by darts continues to be a respectable investment tool. The findings of Dickens and Shelor [2003] supported these conclusions. Using concepts of stochastic dominance they found that the experts' capital gains outperform the darts' but are not superior to any market index; considering total returns including dividends, however, there is no difference between the portfolio of experts and the portfolio of darts. Related issues such as announcement effects and stimulated noise trading in conjunction
with ``investment dartboard columns'' had been discussed by Greene and Smart [1999]. \\
A very detailed and comprehensive study on the performances of portfolios is due to Arnott et al. [2013]. The authors considered various investment strategies combined with different weighting procedures (including upside-down methods) in order to construct portfolios. The resulting performances were then compared with the performances of reference portfolios, in particular, with the performance of the cap-weighted benchmark index. The rich analysis also encompasses simulated random portfolios managed by Malkiel's dart-throwing monkey. It turned out that the dartboard portfolios matched or beat the cap-weighted portfolio in 96 of 100 trials, cf. panel A of exhibit 2 in Arnott et al. [2013]; see also exhibit 1. The authors argue that Malkiel's assessment mentioned above was even too modest because the monkey reliably outperfermed in empirical testing. Moreover, the paradox was found that not only reasonable investment strategies would outperform the cap-weighted benchmark index but also their nonsensical inverted counterparts. These findings give reason to the surmise that investment strategies - sensible or irrational - seem to be more or less arbitrary and eventually yield comparable results. Another profound empirical analysis of the predictive ability of technical trading rules is given by Rink [2023] (see also the literature cited therein). \\
A related phenomenon can be observed apropos certain problems of combinatorical optimization. The situation may roughly be described as follows: if the problem has a small size then the optimum solution - e.g., regarding the computational complexity - would be highly superior to a heuristical or randomly selected solution. As the size of the problem becomes larger, however, the superiority begins to fade, and it may even happen that the efficiencies of the best and the worst solution asymptotically coincide. \\  

The main objective of an investment strategy is of course the attempt to find a portfolio which will perform well in the future. This means that historical data are frequently considered in an explicit or implicit way; furthermore, it should be possible to give an approximate forecast of the future development of assets, at least in connection with the proper selection of weights. \\
It is the purpose of this paper to verify that the situation described above - i.e., there may be no substantial differences between sophisticated and poor approaches - also applies to forecasting of stock returns even under euphemistic model assumptions on the underlying price process. \\

The article is organized
as follows: basic definitions are given in the
next section, existence and explicit representations
for optimal forecasts as well as a detailed
comparison are presented in the third section,
and consequences and the usefulness of optimal
forecasts for practical prediction are discussed 
in the fourth section. The results may loosely
be interpreted as follows:
\begin{itemize}
\item{ \textit{Serious prediction of stock prices
can be achieved only if the best linear unbiased
forecast is used, the planning horizon is small,
and a 'critical relation' is not satisfied.
Otherwise, serious prediction is 
impossible; in particular, the performances of
the trivial forecast and the best measurable
forecast are asymptotically equivalent as the
relative volatility tends to infinity.} }
\item{ \textit{The critical relation states that
the length of the observation interval is too
short in comparison with the squared relative
volatility of the price dynamics. } }
\end{itemize}

\section{Basic assumptions}
In the sequel a Black-scholes model augmented by jumps will 
be considered. However, the proof of the theorem 
formulated in the third section will show that the results actually apply to a 
much broader class of processes (see also the first remark 
at the beginning of the fourth section). \\
The Black-Scholes model postulates that fluctuations
of an asset price $P_t$ at time $t$ follow the equation
\begin{equation}
P_t = P_0 \, \exp( \alpha \, t + \sigma \, W_t), \quad t \geq 0,
\end{equation}
where $P_0$ is the initial price, $\alpha$ stands
for the trend coefficient, $\sigma > 0$ denotes
volatility, and $W_t, t \geq 0$, is a standard
Wiener process with mean values $\E(W_t)=0$ and
covariances $\E(W_s \, W_t)=\min(s,t)$ for $s,t \geq 0$. \\
Consider the following situation: a sample path
of the price process has been observed over the
time interval $[0,T]$, and a potential investor
wishes to estimate the future development of the price
by means of the observed data; for instance, the investor
may estimate the trend coefficient and compute
an appropriate forecast. \\
The Black-Scholes model has been motivated by
diverse arguments. Invoking Donsker's theorem, 
equation (1) can be viewed as a continuous analogue
of the discrete Cox-Rubinstein model. Though
the assumptions of (1) are restrictive the Black-Scholes model
and its extensions are still regarded as useful approximations
and reference points in practice; see, e.g., Ghysels et al. [1996],
Macbeth and Merville[1979], or Leonard and Solt[1990].
Fractional Black-Scholes models have been studied by Bender[2012]
and Xu and Yang[2013], a regime switching version is due to
Mota and Esquivel[2016]. Of course, numerous other models
have been proposed. Well-known classical approaches comprise
ARIMA models (with or without
thresholds), certain diffusion processes, the 
GARCH model and its variants, L\'{e}vy-type
processes, Kalman filtering, etc. Recently discussed methods
with special reference to prediction are based, for instance,
on neural networks (Huang and Huang[2011], Liu and Wang[2012],
Ticknor[2013], Rather et al. [2015]), on fuzzy sets and multivariate
fuzzy time series (Sun et al. [2015]), or on certain nonlinear
relationships between sets of covariates (Scholz et al. [2015]).
Other approaches rely on K-nearest neighbour algorithms (Alkhatib et al. [2013]),
machine learning techniques (Patel et al. [2015]), support
vector regression (Kazem et al. [2013]) or the so-called
CAPS prediction system (Avery et al. [2016]). \\
Discussions on the 'correctness' of stock
price models and references on statistical aspects
may be found on pp.32 of Karatzas and Shreve[1998]
and p. 111 of Musiela and Rutkowski[1997]. In this vein,
see also Bouchaud and Potters[2001], Lauterbach and Schultz[1990]
as well as Amilon[2003]. For some econometric aspects, cf. Campbell et al. [1997]
and the literature given therein. Clearly, this enumeration
is by no means complete.\\

Consider now the problem of optimal prediction. A reasonable
model should reflect possible jumps, cf. Ball and Tourus[1985],
Jorion[1988], or Kou[2002]. Tests indicating the presence of jumps
have been developed by Ait-Sahalia and Jacod[2009]. As the original 
Black-Scholes approach ignores jumps, an extended version is to be
discussed: the price process is assumed to follow
a geometric Brownian motion
\[
 P_t = P_0 \, \exp( \alpha \, t + \sigma \, W_t
       + J_t), \quad t \geq 0
\]
being supplemented by possible jumps $J_t$.
Equivalently, the stock returns $p_t = 
\log(P_t/P_0)$ satisfy the relation
\begin{equation}
 p_t = \alpha \, t + \sigma \, W_t + J_t, \quad t \geq 0.
\end{equation}
The quantities $P_0, \alpha, \sigma$ and $W_t$
are defined as above while $J_t, t \geq 0$, is 
a compound Poisson process, i.e., $J_t = X_1 +
X_2 + \cdots + X_{N_t}$ stands for a random sum
where $N_t, t \geq 0$, denotes a homogeneous
Poisson process with parameter $\lambda > 0$ and
$X_k, k \geq 1$, are independent and identically
distributed random variables with expectations
$\E(X_k)=\nu$ and finite variances $\mathsf{Var}(X_k)
= \tau^2, 0 \leq \tau^2 < \infty$. Compound Poisson
processes are frequently used to describe exogeneous
shocks, particularly in the field of risk management,
cf. Karlin and Taylor[1981] or Schmidt[1996]. \\
All random variables are viewed as real-valued
Borel measurable mappings defined on an underlying
probability space $(\Omega, \A, P)$, and equivalent
random variables coinciding wih probability one
are always identified. Moreover, the Wiener process
$W_t, t \geq 0$, the homogeneous Poisson process
$N_t, t \geq 0$, and the family $X_k, k \geq 1$, 
are assumed to be stochastically independent processes. \\
Setting 
\begin{equation}
\beta = \alpha + \lambda \, \nu \quad \mbox{and} \quad
\mu = \sqrt{\sigma^2 + \lambda(\nu^2 + \tau^2)}
\end{equation}
the first and second moments of the returns are
verified to be 
\begin{equation}
\E(p_t) = \beta \, t \quad \mbox{and} \quad
 \E(p_s \, p_t) = \beta^2 st + \mu^2 \min(s,t)
 \quad \mbox{for} \quad s,t \geq 0.
\end{equation}
$\beta$ can be interpreted as an adjusted trend
coefficient, and both the original volatility
$\sigma$ of the 'normal' market and the 'virtual'
volatility $\sqrt{\lambda(\nu^2 + \tau^2)}$
caused by additional exogeneous shocks yield the
total volatility $\mu > 0$. In the
absence of shocks - i.e., in the special case
$\nu = \tau^2 = 0$ - the conventional Black-Scholes
model (1) is recovered with $\beta = \alpha$
and $\mu = \sigma$. \\
One might argue that there are different types
of shocks having different distributions; at least 
one should distinguish between 'good news' and
'bad news'. Consequently, one should consider
several compound Poisson processes instead of
a single one. A single process, however, is
no restriction because the superposition of 
independent compound Poisson is again a compound
process. \\     

\section{Optimal prediction of stock returns}
\subsection{Optimal prediction of $\bE$ }
Suppose that a sample path of returns $p_t, t \geq 0$,
has been observed over the fixed time interval
$[0,T]$ and let $S > T$ be a prescribed time point.
All parameters $\alpha, \sigma, \lambda, \nu, \tau^2$ are assumed
to be unknown which is the usual situation encountered in practice. The aim is to
specify an 'optimal' forecast of $p_S$ or, more
generally, of $\bE$ where $\B \subset \A$ stands
for a sub-$\sigma$-algebra of $\A$. \\
Consider the Hilbert space $L_2(\Omega, \A, P)$
of real-valued square integrable random variables
being equipped with the usual scalar product. 
There exist at least three well-known criteria
how to choose a best possible element
$Z \in L_2(\Omega, \A, P)$ minimizing the distance
\[
 \Delta(Z) = \E \left(\, (\bE - Z)^2 \,\right)
\]
subject to reasonable side conditions. Let
$\F_t, t \geq 0$, be the canonical filtration
of the process of returns, in other words, $\F_t$
is the smallest sub-$\sigma$-algebra of $\A$ such 
that all returns $p_r, 0 \leq r \leq t$, are
measurable with respect to $\F_t$. Setting
\begin{eqnarray*}
 M_1(T) & = & \{ Z \in L_2(\Omega, \A, P):
   Z \; \mbox{is measurable with respect to} \; \F_T \}, \\
 M_2(T) & = & \{ Z \in L_2(\Omega, \A, P): Z \;
      \mbox{has the representation} \;
      Z = \sum_{i=1}^n \, c_i \, p_{t_i} \\
   &  & \mbox{for some integer} \; n \geq 1, 
\mbox{reals} \; c_i \; \mbox{and time points} \;
                    t_i \in [0,T] \, \}, \\
 M_3(T) & = & \{ Z \in L_2(\Omega, \A, P):
     \E(Z) = \E ( \, \bE \, ) = \beta \, S \, \}
\end{eqnarray*} 
a best measurable forecast $p^\ast_S \in M_1(T)$, 
a best linear forecast $\widetilde{p_S} \in
\overline{M_2(T)}$ and a best linear unbiased
forecast $\widehat{p_S} \in \overline{M_2(T)}
 \cap M_3(T)$ are characterized as solutions
of the respective minimization problems 
\begin{eqnarray*}
 \Delta(p^\ast_S) & \leq & \Delta(Z) \quad 
   \mbox{for all} \; Z \in M_1(T), \\     
 \Delta(\widetilde{p_S}) & \leq & \Delta(Z) \quad 
   \mbox{for all} \; Z \in \overline{M_2(T)} , \\     
 \Delta(\widehat{p_S}) & \leq & \Delta(Z) \quad 
 \mbox{for all} \; Z \in \overline{M_2(T)} \cap M_3(T) .     
\end{eqnarray*}
(Here, $\overline{M_2(T)}$ is the closure of 
$M_2(T)$). In the sequel the symbol
\[
   \gamma = \frac{\mu}{\beta} 
\]
will stand for the relative volatility of the
price process provided the trend coefficient
$\beta$ is nonzero. \\

\textbf{Theorem.} \textit{Suppose $\F_T \subset
\B$ and $\beta \neq 0$. Then the best measurable
forecast $p^\ast_S$, the best linear forecast
$\widetilde{p_S}$ and the best linear unbiased forecast
$\widehat{p_S}$ are given by
\begin{eqnarray}
   p^\ast_S  & = & p_T + \beta \, (S-T) , \\
 \widetilde{p_S} & = & \frac{S+\gamma^2}{T+\gamma^2} \; p_T, \\
 \widehat{p_S}   & = & \frac{S}{T} \; p_T .
\end{eqnarray}
These forecasts are unique, and their respective
mean square errors admit the representations}
\begin{eqnarray}
 \Delta(p^\ast_S) & = & \E \left( (\bE)^2 \right) 
  - (\beta^2 \, S^2 + \mu^2 \, S) 
  + \mu^2 (S-T), \\
 \Delta(\widetilde{p_S}) & = & \E \left( (\bE)^2 \right) 
  - (\beta^2 \, S^2 + \mu^2 \, S) 
  + \mu^2 (S-T) \, \frac{S+\gamma^2}{T+\gamma^2}, \\
 \Delta(\widehat{p_S}) & = & \E \left( (\bE)^2 \right) 
  - (\beta^2 \, S^2 + \mu^2 \, S) 
  + \mu^2 (S-T) \, \frac{S}{T}. 
\end{eqnarray}

\textbf{Proof.} \\
(i) Since the Poisson process $N_t, t \geq 0$, has independent
increments so have the processes $J_t, t \geq 0$, 
and $p_t, t \geq 0$. In particular, the centered
process $p_t - \E(p_t), t \geq 0$, is a martingale
with respect to the canonical filtration.
Hence (5) follows from
\[
p^\ast_S = \E( \, \bE | \F_T \, ) = \E( p_S | \F_T )
         = p_T + \beta (S-T).
\]
Clearly, $p^\ast_S$ is uniquely determined. \\
(ii) In order to show (7) consider an arbitrary
random variable $Z \in M_2(T) \cap M_3(T)$. $Z$
is expressible as a sum $Z = \sum_{i=1}^n \,
c_i \, p_{t_i}$ with time points $t_i \in [0,T]$
and real coefficients $c_i$ where $\E(Z)=\beta\,S$
and $\beta\neq0$ imply $\sum_{i=1}^n \,c_i\,t_i = S$.
$\widehat{p_S} = (S/T)\cdot p_T$ is measurable with
respect to $\B$ and satisfies
\begin{eqnarray*}
 \lefteqn{\E( \, (\bE - \widehat{p_S})(Z - \widehat{p_S})\,) = } \\
 & = & \E( \, \bE \cdot Z \,) - \E(\widehat{p_S} \cdot Z) -
 \E( \, \bE \cdot \widehat{p_S}) + \E(\widehat{p_S}^2) \\
 & = & \E( \, (p_S-\widehat{p_S})(Z-\widehat{p_S}) \, ) \\
 & = & \sum_{i=1}^n \, c_i \, \E(p_S \, p_{t_i}) -
       \sum_{i=1}^n \, c_i \, \frac{S}{T} \, \E(p_T \, p_{t_i}) -
       \frac{S}{T} \, \E(p_S \, p_T) + \frac{S^2}{T^2} \, \E(p_T^2) 
\end{eqnarray*}
whence it follows that
\begin{eqnarray*}
 \lefteqn{\E( \, (\bE - \widehat{p_S})(Z - \widehat{p_S})\,) = } \\
 & = & \sum_{i=1}^n \, c_i \, (\beta^2 \, S \, t_i + \mu^2 \, t_i) \, -
       \sum_{i=1}^n \, c_i \, \frac{S}{T} \, (\beta^2 \, T \, t_i + \mu^2 \, t_i) \\
 &  &  \mbox{} - \frac{S}{T} (\beta^2 \, ST + \mu^2 \, T)
       + \frac{S^2}{T^2} (\beta^2 \, T^2 + \mu^2 \, T) \\ 
 & = & 0.
\end{eqnarray*}
If $Z$ lies in $\overline{M_2(T)} \cap M_3(T)$, choose a sequence
$Z_n \in M_2(T)$ with the property 
$\lim_{n \to \infty} \, \E( \, (Z-Z_n)^2 \, ) = 0$.
According to $ \lim_{n \to \infty} \,
\E(Z_n) = \E(Z) = \beta \, S$ and $\beta \neq 0$
there exists an index $n_0$ with $\E(Z_n) \neq 0$
for $n \geq n_0$. Setting $Z_n' = (\, \beta \, S/\E(Z_n) \,) \cdot Z_n$
one finds $Z_n' \in M_2(T) \cap M_3(T)$ for $n \geq n_0$,
and the continuity of the scalar product in conjunction
with $\lim_{n \to \infty} \, \E( \, (Z-Z_n')^2 \, )
= 0$ ensures 
\[
\E( \, (\bE - \widehat{p_S})(Z - \widehat{p_S})\,) = 0.
\]
$\widehat{p_S}$ is therefore a best linear unbiased 
forecast, and it is unique because the subset
$\overline{M_2(T)} \cap M_3(T)$ is closed and convex.
(6) is verified similarly. \\
(iii) It suffices to calculate $\Delta(\widehat{p_S})$.
One finds
\begin{multline*}
\Delta(\widehat{p_S}) = \\
\E\left( \, (\, \bE - \E(p_S) \,)^2 + 
 2(\, \bE - \E(p_S) \,)(\, \E(p_S) - \widehat{p_S} \,) +
 (\, \E(p_S) - \widehat{p_S} \,)^2 \, \right),
\end{multline*}
and (10) follows from
\[
\E\left( \, (\, \bE - \E(p_S) \,)^2  \,\right)
 = \E\left(\, (\bE)^2 \,\right) - \beta^2 S^2,
\]
\[
\E\left(\,
(\, \bE - \E(p_S) \,)(\, \E(p_S) - \widehat{p_S} \,) \,\right)
= - \E(p_S \, \widehat{p_S}) + \E(p_S)\,\E(\widehat{p_S}) = - \mu^2 \, S
\]
as well as
\[
\E\left(\, 
 (\, \E(p_S) - \widehat{p_S} \,)^2 \,\right) 
 = \mu^2 \, \frac{S^2}{T}.  
\]
\hfill \begin{huge} $\bullet$ \end{huge}       

Relation
\[ 
\beta^2 S^2 = (\, \E(p_S) \,)^2 \leq \E(\,
 (\bE)^2 \,) \leq \E(p^2_S) = \beta^2 S^2 + \mu^2 S
\]
guarantees that the difference
$\E(\, (\bE)^2 \,) - (\beta^2 S^2 + \mu^2 S)$
in (8) - (10) satisfies
\[
 - \mu^2 S \leq \E(\, (\bE)^2 \,) - (\beta^2 S^2 + \mu^2 S) \leq 0.
\]
If inclusion $\F_T \subset \B$ in Theorem 1 is 
replaced by the stronger requirement $\F_S \subset \B$
then the difference in (8) - (10) can be dropped because of
\[ 
\E(\, (\bE)^2 \,) - (\beta^2 S^2 + \mu^2 S) = 0.
\]

The theorem assumes $\beta \neq 0$ but part (i)
of the above proof also applies to $\beta = 0$.
Since $p^\ast_S = p_T$ is then an element of
$\overline{M_2(T)} \cap M_3(T)$ all forecasts coincide:
one obtains
\[
p^\ast_S = \widetilde{p_S} = \widehat{p_S} = p_T
 \quad \mbox{for} \quad \beta = 0.
\]  

\subsection{Comparison between forecasts}
A comparison is to be drawn between the optimal
forecasts discussed in subsection 3.1.
The trivial forecast $p^\circ_S = p_T$ will be
considered also and - for simplicity - attention
is restricted to the sub-$\sigma$-algebra $\B = \A$
which yields $\bE = p_S$. \\
As mentioned above, $\beta = 0$ implies that the
optimal forecasts coincide with the trivial forecast. 
The case $\beta \neq 0$ will therefore be assumed in the
sequel. Table 1 contains both the absolute mean square error
$\Delta(Z)$ as well as the relative performance  
\[
 \delta(Z) = \frac{\Delta(p^\ast_S)}{\Delta(Z)}
\]
with reference to the best measurable forecast
for each $Z \in \{ p^\ast_S, \widetilde{p_S}, 
\widehat{p_S}, p^\circ_S\}$. \\
One finds
\[
\Delta(p^\ast_S) < \Delta(\widetilde{p_S}) <
\min \left( \Delta(\widehat{p_S}), \Delta(p^\circ_S) \right)
\]
with
$\Delta(\widehat{p_S}) < \Delta(p^\circ_S)$ for $T > \gamma^2$,
$\Delta(\widehat{p_S}) = \Delta(p^\circ_S)$ for $T = \gamma^2$
and
$\Delta(\widehat{p_S}) > \Delta(p^\circ_S)$ for $T < \gamma^2$. \\
Of course, the best measurable forecast is always better 
than the best linear forecast which in turn is
always superior to both the best linear unbiased
and the trivial forecast. The best linear unbiased
forecast is worse than the trivial one if and
only if the relation $T < \gamma^2$ holds. \\    
With regard to the planning horizon $S$ the mean
square error of the best measurable forecast is
of order $\mathcal{O}(S)$ as $S \to \infty$
while the order $\mathcal{O}(S^2)$ applies to 
the other forecasts. \\
It is also informative to consider the dependence
of relative performances upon relative volatility
where $T, S$ are held fixed. Since relative
performances are even functions it suffices
to discuss positive values of $\gamma$.\\
$\delta(p^\ast_S)=1$ is of course constant; the
relative performance $\delta(\widehat{p_S}) = T/S$ of
the best linear unbiased forecast is also constant,
and the relative performance $\delta(\widetilde{p_S})$
of the best linear forecast always lies between
$T/S$ and $1$. More precisely, $\delta(\widetilde{p_S})$
is strictly increasing with
\[
 \lim_{\gamma \to 0^+} \, 
 \delta(\widetilde{p_S}) = \frac{T}{S} \quad \mbox{and}
 \quad \lim_{\gamma \to \infty} \, 
 \delta(\widetilde{p_S}) = 1.
\]
$\delta(p^\circ_S)$ is strictly increasing as well
and has the properties
\[
 \lim_{\gamma \to 0^+} \,
 \delta(p^\circ_S) = 0 \quad \mbox{and} \quad
 \lim_{\gamma \to \infty} \,
 \delta(p^\circ_S) = 1
\]
showing that the trivial forecast has the same
asymptotic performance as the best measurable
forecast while $\delta(\widehat{p_S}) = T/S < 1$
remains constant. \\

\newpage

\begin{center}

\textit{Table 1: Comparison between forecasts
$p^\ast_S$ (best measurable),  $\widetilde{p_S}$ (best linear),
$\widehat{p_S}$ (best linear unbiased), and $p^\circ_S$ (trivial)} \\

\vspace{0.3cm}

\begin{tabular}{|c|c|c|c|}
\hline
 Fore- & Formula & Mean square & Relative \\
  cast &         &   error     & performance \\ \hline\hline
 & & & \\  
 $p^\ast_S$ & $p_T + \beta (S-T)$ &
 $\mu^2(S-T)$  & $1$ \\ 
 & & &  \\ \hline
 & & &  \\
 $\widetilde{p_S}$ & $p_T \cdot (S+\gamma^2)/(T+\gamma^2)$ &
 $\mu^2 (S-T)(S+\gamma^2)/(T+\gamma^2)$ &
 $(T+\gamma^2)/(S+\gamma^2)$ \\ 
 & & &  \\ \hline
 & & &  \\
 $\widehat{p_S}$ & $p_T \cdot S/T $ &
 $\mu^2 (S-T)\, S/T$ & $T/S$ \\ 
 & & &  \\ \hline
 & & &  \\ 
 $p^\circ_S$ & $p_T$ & $\mu^2 (S-T) \left( 1 + (S-T)/ \gamma^2 \right)$ &
 $\gamma^2 / (\gamma^2 + S - T)$ \\ 
 & & &  \\ \hline 
\end{tabular}

\end{center}

\subsection{The critical relation}
A direct comparison between the best linear
unbiased forecast and the trivial forecast reveals 
that the inequality $T < \gamma^2$ may be
viewed as a 'critical relation' with
respect to the prediction of future returns by
means of $\widehat{p_S}$.
For if $T$ is too small - i.e., if the length
of the observation interval remains under the
squared relative volatility - then the best linear
unbiased forecast is even worse than the trivial
forecast and becomes highly unreliable. The critical relation
has an essential practical consequence. Predicting
the future development of a given stock price
with relative volatility $\gamma$ an investor 
should therefore avoid $\widehat{p_S}$ on condition
that $T$ does not exceed $\gamma^2$. The squared relative
volatility may therefore be interpreted as a 
'critical time'. Conversely, if the observation
interval $[0,T]$ is given then the usage of 
$\widehat{p_S}$ is particularly risky for
stock prices with $|\gamma| > \sqrt{T}$; 
$\sqrt{T}$ plays the role of a 'critical
relative volatility'. \\
On condition that $T$ is smaller than $\gamma^2$
an investor must trust in the trivial forecast
or refrain from purchasing. This situation poses
the problem of checking the critical relation
$T < \gamma^2$ before making a decision. However,
$\gamma$ depends upon $\mu$ and $\beta$, and an
$L_2$-consistent best linear unbiased estimator of 
$\beta$ generally does not exist for fixed $T$, even in the
absence of exogeneous shocks. (To see this, consider
the special case $\nu = \tau^2 = 0$ characterizing
the absence of shocks. The process
of returns becomes $p_t = \alpha \, t + \sigma \, W_t, 
t \geq 0$, where the linear trend function $f(t) = t$
lies in the kernel reproducing Hilbert space 
associated with the covariance kernel of $\sigma \, W_t$.
Furthermore, the resulting maximum likelihood estimator
$p_T /T$ turns out to be the best linear unbiased estimator of
$\alpha$ with respect to the fixed observation
interval $[0,T]$, and the minimum variance is
$\sigma^2 /T$. See Cambanis[1985] for details on
optimal estimation of trend coefficients in
continuous-time regression models with
correlated errors.)  \\

\subsection{A numerical example}
As an illustration, the relative performances as functions of the
relative volatility $\gamma$ are shown in Figure 1 and Figure 2 for $T=6$ months
and $S=9$ months. This corresponds to an investor's intention to predict
returns one quarter in advance after having observed the price dynamics
over two quarters. The critical relative volatility takes the 
value $\sqrt{6} \approx 2.449$; this means that the trivial forecast 
is better than the best linear unbiased forecast if and only if $|\gamma| > \sqrt{6}$
holds. Figure 1 exhibits the relative performances for small and moderate values of
$\gamma$ ($\gamma \leq 5$); graphically, the curve of the trivial forecast intersects the curve of the best linear unbiased forecast if the argument is equal to $\sqrt{6}$. Relative 
performances for larger values of $\gamma$ ($\gamma > 5$) are given in Figure 2
illustrating that the best linear unbiased forecast becomes poor while the trivial forecast is
asymptotically on a par even with the best possible forecast. \\

\section{Conclusions}
Conclusions always depend upon underlying assumptions;
in this paper, the extended Black-Scholes approach (2)
has been used to describe the price fluctuations. An inspection 
of the proof of the above theorem shows, however, that its assertions 
- and hence the following conclusions - are by
no means restricted to this special approach. Essentially, only 
the independence of increments and properties of the first and 
second moments are required whereas assumptions on distributions
have nowhere been used. Results analogous to the theorem 
can therefore be formulated for more general price processe with different 
trend functions and modified covariance kernels. \\  
Firstly, the advantages of the optimal forecasts
are to be mentioned: they are working in the presence
of exogeneous shocks, and they require only knowledge
of the last return $p_T = \log(\, P_T/P_0 \,)$.
The latter property entails that initial price $P_0$ and
final price $P_T$ are already sufficient with
regard to prediction. Hence one can conclude:  
\begin{itemize}
\item{ \textit{An investor needs to observe both the initial
and the final price; the chart history in the open interval
$(0,T)$, however, is totally irrelevant.} }
\end{itemize}  
Consider the performances of the forecasts. The
best measurable forecast would be satisfactory
because its mean square error is of order $\mathcal{O}(S)$.
Due to the fact that $\beta$ is unknown, however,
$p^\ast_S = p_T + \beta (S-T)$ is practically worthless,
and replacing $\beta$ by a 'reasonable' estimator
does not remedy the situation.
For example, if an arbitrary linear unbiased
estimator is substituted for $\beta$ then the 
resulting forecast cannot be better than $\widehat{p_S}$;
furthermore, recall that an $L_2$-consistent
best linear unbiased estimator of $\beta$ does
not exist in general. In the same vein, the
best linear forecast even depends upon both
$\beta$ and $\mu$. $p^{\ast}_S$ and $\widetilde{p_S}$
are therefore only of theoretical interest. \\
The best linear unbiased forecast $\widehat{p_S}$
is left but its application may be a malicious
alternative. The main problem occurs if the critical
relation $T < \gamma^2$ is satisfied which implies
that the best linear unbiased forecast is even
worse than the trivial forecast. Intuitively,
this situation admits the following interpretation:
if a stock price has a large relative volatility
 - i.e., if the trend coefficient $\beta$ is 
small in comparison with the total volatility
$\mu$ - and if, in addition, the observation interval 
$[0,T]$ is short then a precise differentiation
between trend and noise is impossible. Consequently,
the extrapolation $\widehat{p_S} = (S/T) \cdot p_T$
will be highly susceptible to non-systematic
random fluctuations while the trivial forecast
$p^\circ_S = p_T$ will be less affected. In particular,
$\widehat{p_S}$ will drastically overestimate
the actual return $p_S$ provided $p_T$ happens
to be atypically large. \\
The critical relation holds true especially for
sufficiently small values of $|\beta|$ corresponding
to a stock price which is 'moving sideways'. Note
that $\widehat{p_S}$ implicitly depends upon
$\beta$ insofar as $\beta \neq 0$ has been assumed
in the theorem. If $\beta$ is exactly equal to $0$
then $\widehat{p_S}$ coincides with the trivial
forecast. \\
These considerations lead to the following
conclusion. 
\begin{itemize}
\item{ \textit{Prediction of stock prices is 
extremely risky if the critical relation is satisfied.} }
\end{itemize}
Suppose that the critical relation does not hold.
Being of order $\mathcal{O}(S^2)$ the mean square
error of $\widehat{p_S}$ might then be acceptable
for short planning horizons but a substantial
error is obtained for medium-sized and large
values of $S$. This is also illustrated by
the constant relative performance $\delta(\widehat{p_S})
= T/S$ which rapidly becomes poor as $S$ increases.
These facts imply:
\begin{itemize}
\item{ \textit{Serious forecasts can be
achieved only in case of sufficiently small
relative volatilities in conjunction with short
planning horizons.} }
\end{itemize}

Practical experience and activities such as the Wall Street Journal's dartboard contest 
show that prediction of stock prices is more or less fallacious. 
This dubiety is confirmed by the conclusions mentioned above - with the
exception that a reasonable forecast may be obtained in case of
sufficiently small relative volatilities. However, this statement
has been derived from a model assuming a constant volatility.
Though a constant volatility is often regarded as useful approximation
and reference point, volatility is in fact a stochastic process, 
cf. Ghysels et al. [1996].
This will result in an increased uncertainty,
in other words: if prediction is already
questionable for constant volatilities then one cannot
expect that dubiety of forecasts will be mitigated by
variable volatilities. \\

\end{large}

\newpage

\section*{References}

\textbf{Ait-Sahalia, Y., Jacod, J. (2009)}:
Testing for Jumps in a Discretely Observed Process.
The Annals of Statistics 37, 184-222. \\ 
\textbf{Alkhatib, K., Najadat, H., Hmeidi, I., Shatnawi, M.K.A. (2013)}:
Stock Price Prediction Using K-Nearest Neighbour Algorithm.
International Journal of Business, Humanities and Technology 3, 32-44. \\
\textbf{Amilon, H. (2003)}:
A Neural Network versus Black-Scholes: A Comparison of
Pricing and Hedging Performances. Journal of Forecasting 22, 317-335. \\
\textbf{Arnott, R.D., Hsu, J., Kalesnik, V., Tindall, P. (2013)}:
The Surprising Alpha From Malkiel's Monkey and Upside-Down Strategies.
The Journal of Portfolio Management 39, 91-105. \\
\textbf{Avery, C.N., Chevalier J.A., Zeckhauser, R.J. (2016)}:
The "Caps" Prediction System and Stock Market Returns.
Review of Finance 20, 1363-1381. \\
\textbf{Ball, C., Tourus, W. (1985)}:
On Jumps in Common Stock Prices and their Impact on Call
Option Pricing. The Journal of Finance 40, 155-173.\\
\textbf{Bender, C. (2012)}:
Simple Arbitrage. The Annals of Applied Probability 22, 2067-2085. \\
\textbf{Bouchaud, J.P., Potters, M. (2001)}:
Welcome to a Non-Black-Scholes World. Quantitative Finance 1, 482-483. \\
\textbf{Cambanis, S. (1985)}: Sampling Designs
for Time Series. In: Time Series in the Time Domain.
Handbook of Statistics, Vol.5, 337-362, North-Holland, 
Amsterdam. \\  
\textbf{Campbell, J.Y., Lo, A.W., MacKinlay, A.C. (1997)}:
The Econometrics of Financial Markets. Princeton
University Press, Princeton. \\
\textbf{Dickens, R., Shelor, R. (2003)}:
Pros Win! Pros Win! ... or Do They? An Analysis of the ``Dartboard'' Contest Using Stochastic Dominance. Applied Financial Economics 13, 573-579. \\
\textbf{Ghysels, E., Harvey, A.C., Renault, E. (1996)}:
Stochastic Volatility. In: Handbook of Statistics, Vol. 14, 
119-191, Elsevier. \\
\textbf{Greene, J., Smart, S. (1999)}: Liquidity Provision and Noise Trading: Evidence from
the `Investment Dartboard Column'. The Journal of Finance 54, 1885-1899. \\
\textbf{Huang, L., Huang, S. (2011)}:
Optimal Forecasting of Option Prices Using Particle Filters
and Neural Networks. Journal of Information and Optimization Sciences 32, 255-276. \\
\textbf{Jorion, P. (1988)}:
On Jump Processes in the Foreign Exchange and Stock Markets.
Review of Financial Studies 1, 259-278. \\
\textbf{Karatzas, I., Shreve, S.E. (1998)}:
Methods of Mathematical Finance. Springer, New York. \\
\textbf{Karlin, S., Taylor, H.W. (1981)}: A
Second Course in Stochastic Processes. Academic
Press, London. \\
\textbf{Kazem, A., Sharifi, E., Hussain F.K. (2013)}:
Support Vector Regression with Chaos-Based Firefly Algorithm
for Stock Market Price Forecasting. Applied Soft Computing 13, 947-958. \\ 
\textbf{Kou, S.G. (2002)}:
A Jump-Diffusion Model for Option Pricing. Management Science 48, 1086-1101. \\
\textbf{Lauterbach, B., Schultz, P. (1990)}:
Pricing Warrants: An Empirical Study of the Black-Scholes Model
and Its Alternatives. The Journal of Finance 45, 1181-1209. \\
\textbf{Leonard, D.C., Solt, M.E. (1990)}:
On Using the Black-Scholes Model to Value Warrants.
Journal of Financial Research 13, 81-92. \\
\textbf{Liu, F., Wang, J. (2012)}:
Fluctuation Prediction of Stock Market Index by Legendre
Neural Network with Random Time Strength Function.
Neurocomputing 83, 12-21. \\
\textbf{Macbeth, J.D., Merville, L.J. (1979)}:
An Empirical Examination of the Black-Scholes Call Option
Pricing Model. The Journal of Finance 34, 1173-1186. \\
\textbf{Malkiel, B. (2007)}:
A Random Walk Down Wall Street. W.W. Norton \& Company, Inc., New York. \\
\textbf{Metcalf, G.E., Malkiel, B.G. ((1994)}: The Experts, the Darts, and the Efficient 
Market Hypothesis. Applied Financial Economics 4, 371-374. \\
\textbf{Mota, P.P., Esquivel, M.L. (2016)}:
Model Selection for Stock Price Data. Journal of Applied Statistics 43, 2977-2987. \\
\textbf{Musiela, M., Rutkowski, M. (1997)}:
Martingale Methods in Financial Modelling. 
Springer, New York. \\
\textbf{Patel, J., Shah, S., Thakkar, P. (2015)}:
Predicting Stock and Stock Price Index Movement Using 
Trend Deterministic Data Preparation and Machine Learning
Techniques. Expert Systems with Applications 42, 259-268. \\
\textbf{Rather, A.M., Agarwal, A., Sastry, V.N. (2015)}:
Recurrent Neural Network and a Hybrid Model for Prediction
of Stock Returns. Expert Systems with Applications 42, 3234-3241. \\
\textbf{Rink, K. (2023)}: The Predictive Ability of Technical Trading Rules: an Empirical
Analysis of Developed and Emerging Equity Markets. Financial Markets and Portfolio Management 37, 403-456. \\
\textbf{Schmidt, K.D. (1996)}: Lectures on Risk
Theory. Teubner, Stuttgart. \\
\textbf{Scholz, M., Nielsen, J.P., Sperlich, S. (2015)}:
Nonparametric Prediction of Stock Returns Based on Yearly Data:
The Long-Term View. Insurance: Mathematics and Economics 65, 143-155. \\
\textbf{Sun, B., Guo, H., Karimi, H.R., Ge, Y., Xiong, S. (2015)}:
Prediction of Stock Index Future Prices Based on Fuzzy Sets
and Multivariate Fuzzy Time Series. Neurocomputing 151, 1528-1536. \\
\textbf{Ticknor, J.L. (2013)}:
A Bayesian Regularized Artificial Neural Network for Stock
Market Forecasting. Expert Systems with Applications 40, 5501-5506. \\
\textbf{Xu, S., Yang, Y. (2013)}:
Fractional Black-Scholes Model and Technical Analysis of
Stock Price. Journal of Applied Mathematics, Article ID 631795, 7 p.  \\

\newpage

\begin{large}
\begin{center}
 \textit{Figure 1: Relative performance against relative volatility}  
\end{center} 
\end{large}
\vspace{0.1cm}
 \begin{tikzpicture}[scale=1.7,domain=0:5]
 \begin{axis}
\addplot[no markers, color=black,style=thick]{1};
\addplot[no markers, color=blue,style=thick]{(6+x^2)/(9+x^2)};
\addplot[no markers, color=green,style=thick]{2/3};
\addplot[no markers, color=red,style=thick]{x^2/(3+x^2)};
\end{axis}

\draw[color=black,style=very thick] (3,2) -- (3.5,2) node[right,color=black]{best measurable};
\draw[color=blue,style=very thick] (3,1.5) -- (3.5,1.5) node[right,color=black]{best linear};
\draw[color=green,style=very thick] (3,1) -- (3.5,1) node[right,color=black]{best linear unbiased};
\draw[color=red,style=very thick] (3,0.5) -- (3.5,0.5) node[right,color=black]{trivial};

\draw  (3.5,-0.75)  node{\begin{large}\textbf{Relative volatility}\end{large}};
\draw  (0,6)  node{\begin{large}\textbf{Relative performance}\end{large}};

\end{tikzpicture}  

\newpage

\begin{large}
\begin{center}
 \textit{Figure 2: Relative performance against relative volatility}  
\end{center} 
\end{large}
\vspace{0.1cm}
 \begin{tikzpicture}[scale=1.7,domain=5:20]
 \begin{axis}
\addplot[no markers, color=black,style=thick]{1};
\addplot[no markers, color=blue,style=thick]{(6+x^2)/(9+x^2)};
\addplot[no markers, color=green,style=thick]{2/3};
\addplot[no markers, color=red,style=thick]{x^2/(3+x^2)};
\end{axis}

\draw[color=black,style=very thick] (3,3) -- (3.5,3) node[right,color=black]{best measurable};
\draw[color=blue,style=very thick] (3,2.5) -- (3.5,2.5) node[right,color=black]{best linear};
\draw[color=green,style=very thick] (3,2) -- (3.5,2) node[right,color=black]{best linear unbiased};
\draw[color=red,style=very thick] (3,1.5) -- (3.5,1.5) node[right,color=black]{trivial};

\draw  (3.5,-0.75)  node{\begin{large}\textbf{Relative volatility}\end{large}};
\draw  (0,6)  node{\begin{large}\textbf{Relative performance}\end{large}};

\end{tikzpicture}  

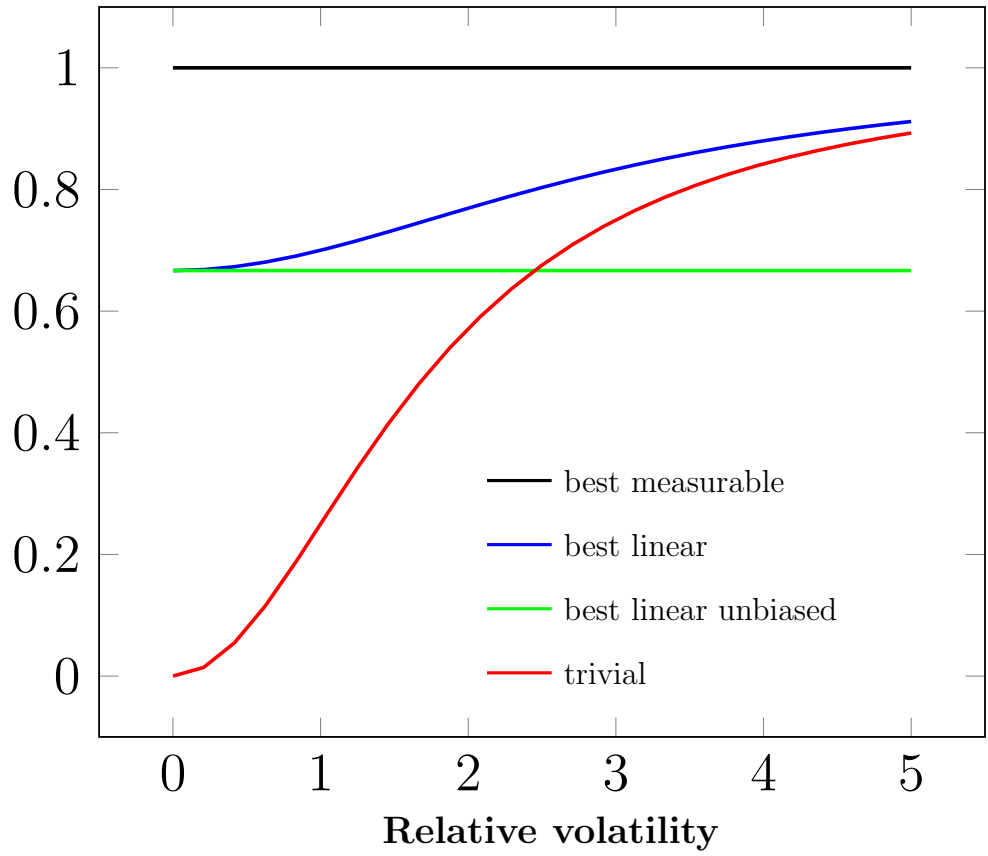
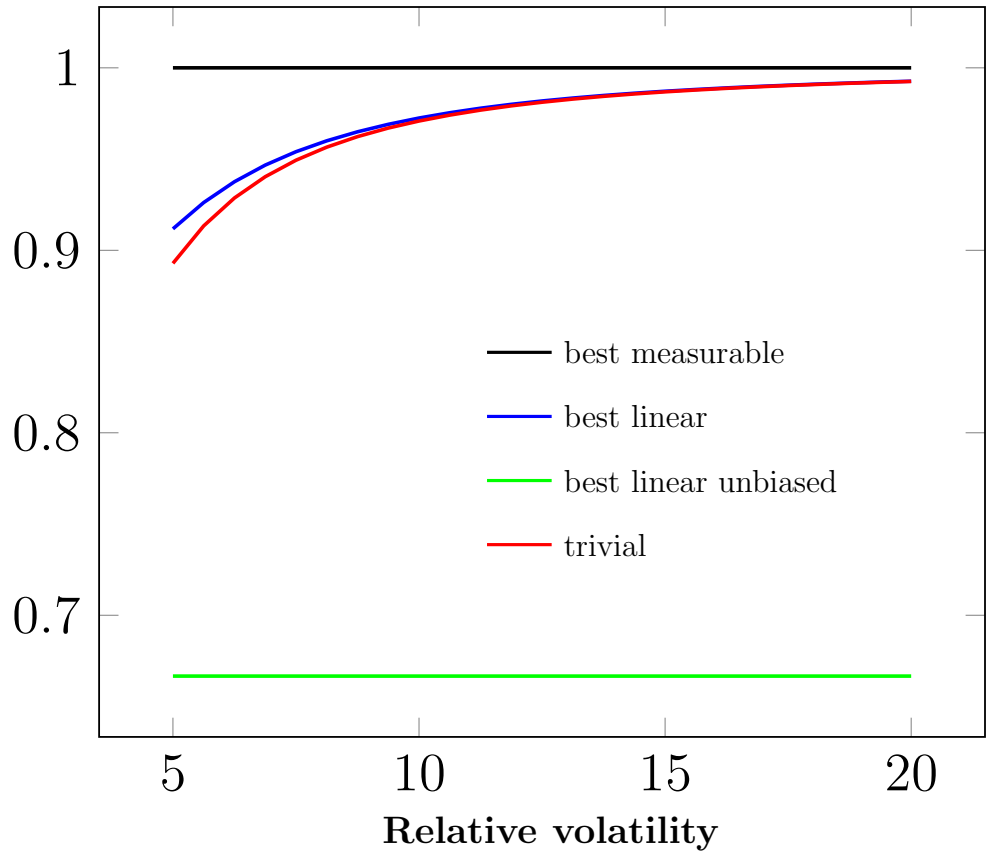
\end{document}